\documentclass[12pt]{iopart}
\begin{document}

\title[The new astrophysical scenario fixed by 
experimental evidences]
{The new astrophysical scenario fixed by the Einstein's 
equivalence principle and gravitational experiments}                                                                        

\author{Rafael A. Vera\dag\footnote[3] {rvera@udec.cl}}

\address{\dag\ Departamento de Física. Facultad de Ciencias 
Físicas y Matemáticas. Universidad de Concepción. Concepción. 
Chile.}

\begin{abstract}
The Einstein's equivalence principle and experiments in which 
bodies and observers are in different G potentials have been 
used to prove that the chain of hypothesis coming from 
assuming the absolute invariability of the bodies, after a 
change of G potential, are wrong. The absolute changes of 
frequencies, masses and lengths of every well-defined part of 
the non-local bodies, compared with the local ones, are 
linearly related to the differences of G potential between the 
non-local bodies and the observer. Such absolute changes are 
independent on the forces within the structure of the bodies. 
The increase of G potential due to universe expansion expands 
bodies in same proportion as distances. Consequently, such 
uniform expansion cannot change the results of any measurement 
of distance or velocity.  The cosmological red shifts and the 
average universe density cannot change with the time, i.e., 
the new universe age is not limited by the Hubble law. The new 
linear G relationships and the unlimited age of the universe 
fix a new astrophysical context that turns out to be 
independent on any cosmological hypothesis. Then the best fit 
of this new scenario with astronomical observations can be 
obtained independently on the chain of hypotheses that are not 
simultaneously consistent with all of the experimental facts. 
\end{abstract}



\maketitle

\section{Introduction}

Current physics and astrophysics are based on direct relationships between
quantities measured by observers located in different gravitational (G)
potentials. This is equivalent to assume that the observers and their
reference standards are absolutely invariable after a change of G potential.
This is the last \textit{classical hypothesis} that is still present in
physics.

For example, in the case of a free fall and stop in a G field, according to
the classical hypothesis, the initial and the final rest masses of a body,
with respect to the observer at the end of the fall, are the same. \textit{\
This is equivalent to rule out the possibility that the body can put on the
energy for the G work}. Thus 
\textit{the Einstein's hypothesis in that the G field puts on the energy for
the G work.}\cite{Einstein} is a direct consequence of assuming that the
classical hypothesis is true.

On the other hand the classical hypothesis is in clear contradiction with
the G time dilation (GTD) experiments made up with standard atomic clocks.
From them, an observer in a fixed G potential \textbf{A }finds that \textit{
a standard atomic clock located at rest in a higher G potential \textbf{B}
has a higher frequency compared with that of his local clock at }\textbf{A}.
The difference of frequency, with respect to the observer at \textbf{A}, is
proportional to the difference of G potential between such clocks\cite
{Hafele} \cite{Vessot}.

Then the positive results of the GTD experiment prove that:

\begin{itemize}
\item  \textit{Some fundamental physical changes have occurred to the
standard clock and its atoms after a change of G potential, compared with
clocks that have not changed of G potential.}

\item  \textit{The classical hypothesis is not consistent with experimental
facts.}
\end{itemize}

\begin{description}
\item  The real (absolute) physical changes occurring to the bodies after a
change of G potential, with respect to any observer that does not change of
potential, have been derived before and below, from theoretical and 
experimental approaches
\cite{Vera1} \ \cite{Vera2}\ \cite{Vera3}\ \cite{Vera4}\ \cite{Vera5}\ 
\cite{Vera6}\ \cite{Vera7}\ \cite{Vera8}\ \cite{Vera9}\ \cite{Vera10}
\end{description}

Since the standard clocks located in different G potentials are physically
different with respect to each other, then most of the current relationships
between quantities measured by observers at rest in different G potential
are inhomogeneous. They have been the source of errors and confusions in the
current literature, from long time ago.

\subsection{Discussion}

-- \textit{It has been argued that the positive results of the GTD
experiments made up with clocks and electromagnetic signals were due to a
presumed frequency change that the photons would have, according to general
relativity (GR), during the trips of radiation traveling between the 
clocks.}

Such argument is wrong for several reasons:

\begin{enumerate}
\item  \textit{The measured time intervals between the signals do not depend
on the frequency of the photons}.

\item  The presumed change of frequency of the photons is not consistent
with ``wave continuity'', which is one of the best-tested properties of
electromagnetic waves. From it, not a single wave can be lost or created
during the trip between the potentials \textbf{B} and \textbf{A. }All of the
waves of a wave train should take the same time to travel between \textbf{B}
and \textbf{A}. Then the net number of waves of a wave train crossing any
static plane, between \textbf{B} and \textbf{A}, should remain unchanged,
i.e., the theoretical frequency of the non-local (NL) waves, with respect to 
any clock of constant frequency, is bounded to be constant.

\item  Such argument is based on a ``vicious circle'' in which the theory of
GR is used to try to justify the classical hypothesis tacitly used in the
same theory, regardless on the fact that the GTD experiments prove just the
opposite.

\item  In the Hafele--Keating experiments, electromagnetic signals were not
used. The clocks readings were directly compared before and after the
experiments.
\end{enumerate}

-- \textit{Most people believes in that the positive results of the GTD
experiments and other gravitational tests would have verified the Einstein's
theory on general relativity }(GR).

This is not true because the Einstein's theory on GR is certainly based on 
the EEP but it is also based on another independent hypothesis on 
\textit{the G field energy. } The last hypothesis, as shown in the above 
example, is equivalent to assume that the classical hypothesis is true. Such 
hypothesis is in clear contradiction with the physical changes observed in 
the GTD experiments by observers that do not change of G potential. Such 
hypothesis is also in contradiction with the fact that the gradient of a 
strictly static G field does not travels altogether with the body. It is 
well known that strictly static forces give up just momentum. They do not 
give up energy. 

\subsection{The formalism fixed by the experimental facts}

\textit{In static conditions}, according to results of GTD experiments, the
frequency of a NL clock in a G potential \textbf{B,}\textit{\
with respect to an observer in a G potential }\textbf{A}, is a function of
the difference of G potential of the NL clock at \textbf{B} with respect to
the observer's clock at \textbf{A}, called\textbf{\ } $\Delta \phi 
_{A}(B\;)$. Notice that the special position or G potential of the observer 
\textbf{A}is stated by a subscript

Assume that $\Delta E_{A}$ is the initial energy given after an upward
impulse to a clock at \textbf{A} and that such clock stops at \textbf{B.}
Assume that a GTD experiment done in the small time interval in which the
clock is static. Then the results of such experiment, with respect to the
observer at \textbf{A}, can be described by:

\begin{equation}
\frac{\nu _{A}(B)-\nu _{A}(A)}{\nu _{A}(A)}=\Delta \phi _{A}(B)=\frac{\Delta
E_{A}}{m_{A}(A)}  \label{1}
\end{equation}

Notice that $\Delta \phi _{A}(B\;)$ is the dimensionless form of the G
potential change in which the mass unit is $1$ joule.

\textit{In non-static cases}, from special relativity, the frequency of the
NL clock at \textbf{B} also depends on its velocity ($V$) with respect to
the observer at \textbf{A}. Thus the symbol $\nu _{A}(V,B)$ has been used in
previous works.

From this result, it is obvious that the reference standards at $A$ and $B$ are
not the same with respect to each other. Then, to relate quantities measured
in different G potentials, all of the quantities must be transformed to a
common unit system based on a reference standard (or a clock\textit{)} in a
well-defined G potential, say \textbf{A}.

In previous works, the G transformations factors have been derived
theoretically, after using a particle model consistent with general
experimental facts\cite{Vera1} \cite{Vera2}. A more direct way has been
done, more recently, after applying the Einstein's equivalence principle
(EEP) to GTD experiments in which bodies and observers are in different G
potentials\cite{Vera9} \cite{Vera10}.

\subsection{Generalization of the Einstein's equivalence principle for non
local cases in G fields}

From the EEP it is found that: when an observer and his measuring system 
change from a rest at the G potential \textbf{A} up to the G potential
\textbf{B} he cannot detect, from local measurements, any local physical
change. This is because \textit{accurate experiments prove that all of the
local ratios within his local measuring system are constants that do not
depend on its position with respect to other systems}\cite{Misner}.

On the other hand, from the GTD experiments it is found that the observer 
at \textbf{A} detect a difference of frequency of the clock at \textbf{B}, 
with respect to the clock at \textbf{A} , which is proportional to the 
difference of G potential between such clocks.

These two well-tested experimental facts can be consistent to each other
only if:

\textit{The absolute values of all of them, the frequencies, the masses and
the lengths of every well-defined particle and standing wave of the same
system change ``linearly'', in just the same the same proportion as the
frequency of the standard clock, after a common change of G potential}%
.\medskip \medskip After use of (\ref{1})

\begin{equation}
\frac{\nu _{A}(B)-\nu _{A}(A)}{\nu _{A}(A)}=\frac{m_{A}(B)-
m_{A}(A)}{m_{A}(A)%
}=\frac{\lambda _{A}(B)-\lambda _{A}(A)}{\lambda _{A}(A)}=\Delta \phi
_{A}(B)=\frac{\Delta E_{A}}{m_{A}(A)}  \label{2}
\end{equation}

From equation (\ref{2}) it is concluded that:

\begin{enumerate}
\item  Some real (absolute) physical changes do occur to any standard body
after a change of G potential, i.e., the classical hypothesis is not true.

\item  Direct relationships between quantities measured by observers at rest
in different G potential are inhomogeneous because they are referred to
standards that are physically different compared to each other.

\item  The field equations fixed by the EEP are strictly linear.

\item  The mass-energy of a NL body at rest with respect to a fixed
observer is not constant. It depends on the difference of G potential
between the body and the observer.
\end{enumerate}

\subsection{Inconsistency of the G field energy hypothesis}

From the second and fifth members of (\ref{2})

\begin{equation}
\Delta m_{A}(B)=\Delta E_{A}(B)  \label{3}
\end{equation}

From (\ref{3}), the G energy comes not from the G field but from the test
body (\textbf{B}). A fraction of its rest mass is transformed into free
energy, or vice versa. Then,

\begin{itemize}
\item  \textit{There is not a real exchange of energy between a static field
and the body.}

\item  \textit{The G red shift phenomena is not due to a change of frequency
of photons. It comes from different eigen-frequencies of atoms located in
different G potentials. }

\item  \textit{Radiation and G fields exchange just momentum like in any
refraction phenomenon. }
\end{itemize}

These results are in clear contradiction with the Einstein's hypothesis on
the energy exchange between bodies or radiation and G fields .

\section{The new physical approach based on a particle model}

\subsection{The particle model}

According to the EEP, any well-defined standing wave of a measuring system
must obey the same inertial and gravitational laws as any other uncharged
particle of the same system. If this were not so, the differences could be
detected from Michelson-Morley experiments made up after changes of velocity
and G potential. Such positive measurements would violate the EEP.

Thus from the EEP it is concluded that \textit{the inertial and
gravitational properties of uncharged particles of a system can be derived
from particle models made up of photons in stationary state within the
same system. }\medskip

In the first steps of this work, such particle model was justified otherwise
from experiments in which matter is transformed into radiation and vice
versa. Then the theoretical properties of uncharged bodies and their G
fields were derived from a particle model made up of some quantum of
electromagnetic radiation confined as standing waves within perfect mirrors.
Its local mass-energy was defined by:

\begin{equation}
E_{A}(A)=m_{A}(A)=nh\nu _{A}(A)  \label{4}
\end{equation}

Thus equation (\ref{2}) was derived, independently on the EEP, from just
general properties of radiation\cite{Vera1} \cite{Vera2}. Then such
primitive approach may be regarded as a clear verification of the more
general form of the EEP.

However, since the EEP is one of the best tested principles of physics, then
the new approach starting from the EEP is more reliable and simpler, as
shown below.

On the other hand, the more detailed deductions done in the first works do
provide a deeper and unified understanding on the nature of the physical
phenomena occurring in the bodies and in the space in G fields. From them it
is simple to verify that this model accounts for basic relationships of 
\textit{special relativity, quantum mechanics and for all of the
conventional gravitational tests}\cite{Vera1} \cite{Vera2} \cite{Vera7} .

\subsection{Theoretical properties of the ``non-local'' space in static
gravitational fields}

\subsubsection{The speed of non-local light}

If a particle model is moved from the observer's potential \textbf{A} up to
a higher G potential \textbf{B} , then the speed of light in the NL model at 
\textbf{B}, compared with the one at \textbf{A}, is fixed by the product of
its frequency and its wavelength with respect to the observer at \textbf{A}.

\begin{equation}
c_{A}(B)=\nu _{A}(B)\lambda _{A}(B)  \label{5}
\end{equation}

From (\ref{5}) and (\ref{2}), the proportional difference of the speed of NL
light at \textbf{B} with respect to the observer at \textbf{A} is:

\begin{equation}
\frac{\Delta c_{A}(B)}{c_{A}(A)}=\frac{\Delta \nu _{A}(B)}{\nu _{A}(A)}+%
\frac{\Delta \lambda _{A}(B)}{\lambda _{A}(A)}=2\Delta \phi _{A}(B)
\label{6}
\end{equation}

From (\ref{6}) it is concluded that the gradient of G potential comes from 
\textit{a gradient of the refraction index of the space with respect to
observers at rest in fixed (invariable) G potentials}. Such gradient
directly accounts for the results of experiments on:\cite{Vera2}

\begin{itemize}
\item  Time delay of radar waves traveling close to the Sun.

\item  Deviation of light traveling close to massive stars (lens effect).
\end{itemize}

From (\ref{6}) it is concluded that \textit{the interaction of an
electromagnetic wave with a G field is a refraction phenomenon produced by a
gradient of the speed of NL light.} Since the refraction phenomenon does not
change the frequency (color or energy) of photons, then the experiments 1.
and 2. are also two independent ways to prove that \textit{there is not a
real exchange of energy between the G field and radiation}.

Then either from (\ref{6}) or from just wave continuity it is concluded 
that:

\begin{itemize}
\item  \textit{During the trip of an electromagnetic wave in a G field, its
frequency with respect to an observer (or a clock) in a fixed potential,
remains constant.} (Conservation law for the frequency of NL radiation with
respect to strictly invariable clocks)

\item  \textit{There is not a true energy exchange between the G field and
radiation.}

\item  \textit{The G red shift phenomenon does not occur during the trip of
the radiation. It comes from a difference in the natural (eigen) frequencies
of the standard atoms located at rest in different G potentials}
\end{itemize}

From (\ref{6}) and (\ref{2}) it is also inferred that the eigen values of
the frequencies, masses and lengths of the NL bodies at rest in a G field,
with respect to a fixed observer, depend on the square root of the speed of
NL light with respect to such observer.

\subsection{Gravitational properties of space from matter distribution in 
the universe}

When all of the uncharged particles of the universe are emulated by particle
models it is found that the universe is a sea of wavelets that interfere
constructively only in the sites where photons and particles are. In the
space between them, the wavelets would interfere destructively, with random
phases, i.e., the probability for the existence of energy in such place is
null. This would account for \textit{the lack of energy of the G field}.

Then the G properties of the empty space can depend only on the \textit{net
space perturbation rate} produced by all of the wavelet with random phases
crossing it. Thus the space properties at some arbitrary position $r^{i}$
should be proportional to the sum of the contributions of all of the
wavelets crossing the same space. Such contributions must be proportional to 
\textit{the products of the frequency and the amplitude of the wavelets
actually crossing the space}.

During the wavelet trips, their frequencies should be attenuated by the same 
\textit{cosmological red shift of light}, according to $d\nu /\nu =dr/R$ in
which $R$ is the Hubble radius. Then, for simplicity, the net wavelet
perturbation rate of the space can be defined by:

\begin{equation}
\sum_{j=1}^{\infty }\frac{\nu ^{j}}{r^{ij}}\exp \left[ \frac{r^{ij}}{R}%
\right] \propto \sum_{j=1}^{\infty }\frac{m^{j}}{r^{ij}}\exp \left[ \frac{%
r^{ij}}{R}\right] =w(r^{i})  \label{7}
\end{equation}

For a uniform density ($\rho $) of the universe, after integration of (\ref
{7}), $w(r)=4\pi \rho R^{2}.$

It is simple to verify that the best correspondence of (\ref{7}) and 
(\ref{2}
) with the experimental facts occurs when eigen frequencies of the particles
are in a sort of equilibrium with the perturbation rate of the space in
which they are located, i.e., when:

\begin{equation}
w(r)\nu (r)=Constant  \label{8}
\end{equation}

From (\ref{2}) and (\ref{8}),

\begin{equation}
\Delta \phi (r)=\frac{\Delta \nu (r)}{\nu (r)}=
-\frac{\Delta w(r)}{w(r)}  \label{9}
\end{equation}

In a central field, $\Delta w(r)$ $=-M\Delta r/r^{2}$, which is extremely
small compared with $w(r).$ Then the factor $1/w(r)$, called $G(r)$, is
nearly constant. Since the mass unit used here is 1 joule, then this new
constant is related to the current constant $\mathbf{G}$ by:

\begin{equation}
\frac{1}{w(r)}\approx \frac{1}{4\pi \rho R^{2}}=G(r)\approx \mathbf{G}c^{-4}
\label{10}
\end{equation}

In weak fields, equations (\ref{9}), (\ref{7}) and (\ref{10}) are consistent
with the Newton's law and with the general tests for G theories.

The universe density derived from (\ref{10}) is about 30 times the average
density of luminous matter of the universe. This one is consistent with the
proportions of \textit{dark matter} estimated in some clusters.

\subsection{The linear black hole}

The new G relationships fixed by the EEP are strictly ``\textit{linear''}.
Then the theoretical properties of the new kind of ``linear'' black hole
(LBH) are different from those of the ``non-linear'' black hole of GR \cite
{Vera2} \cite{Vera7}.

The LBH should be like a giant atomic nucleus, or neutron star, that obeys
the nuclear physical laws. From (\ref{6}), the high gradient of the NL
refraction index that should exist around it must prevent, after critical
reflection, the escape of photons and high-speed particles.

Consequently, the LBH must absorb all kind of radiation and emit almost
nothing. The average NL mass-energy of its neutrons must increase with the
time. When such mass becomes higher than the one in free state, the LBH can 
explode after producing a rather spherical volume of gas.

The nucleons escaping collectively from a LBH would decay, mainly, into new
hydrogen gas (and antineutrinos) rather free of metals. This makes a radical
difference with current star explosions. Of course, other kinds of nuclear
decays may eventually occur. Older bodies
orbiting around the LBH may capture the new gas. They may become planets or
stars whose masses depend on their original masses.

\section{Gravitational expansion matter during universe expansion}

From equation (\ref{2}), there is a phenomenon of ``\textit{gravitational
expansion'' of the bodies after an increase of G potential}. \textit{Such
expansion cannot be prevented by the internal forces within the structure of
the bodies because equation (\ref{2}) comes from the EEP and the GTD
experiments, which are independent on the structure of the bodies.}

Then the general increase of G potential occurring during universe expansion
must produce a G expansion of all of the bodies in it.

After some time interval $\Delta t$, all of the distances should increase in
the same proportion. Then the general increase of G potential turns out to
be just equal to the proportional increase of distance between any generic
particle like $i$ and $k$, called $r^{ik}$ \cite{Vera13}.

\begin{equation}
\Delta \phi =\frac{\Delta r^{ik}}{r^{ik}}=H\Delta t  \label{11}
\end{equation}

From (\ref{2}) and (\ref{11}), the G expansion of a standard rod is given 
by:

\begin{equation}
\frac{\Delta \lambda }{\lambda }=\Delta \phi =\frac{\Delta r^{ik}}{r^{ik}}%
=H\Delta t  \label{12}
\end{equation}

From (\ref{12}) it is concluded that

\begin{enumerate}
\item  According to the EEP, only an absolute kind of universe expansion 
can exist. In it, bodies and distances should expand in just the same
proportion. Then it would be impossible to find a reference frame that 
does not expand in just the same proportion.

\item  Such expansion cannot not produce relative changes of distances,
velocities and cosmological red shifts. The general conservation laws of
physics would not violated.

\item  The average universe should look the same, in any time, i.e., the
universe age, after the cosmological period, is indefinite.
\end{enumerate}

The same result is obtained after emulating particles by particle models.
The positions of the particles would be fixed by constructive interference
of a sea of wavelets interlacing them. After Doppler shift, all of the
wavelets would expand in the same proportion without changing the net number
of wavelets and wavelengths between the bodies, i.e., without changing any
distance measurement. Then the cosmological red shifts cannot change with 
the time.

Notice that the use of particle models and wavelets opens another
alternative explanation for the cosmological red shift. Which is that there
is a wavelet red shift proportional to the distances, most probably produced
by diffraction, i.e., by some kind of unknown kind of \textit{%
wavelet-wavelet interaction}. Statistically, to the contrary of photon
scattering, the new alternative would not produce appreciable widening of
the spectrum lines. It seems to be equivalent to that of a universe 
expansion.

\section{The new astrophysical scenario fixed by the EEP}

From the new properties of the black holes and of the unlimited age of 
the universe, it is inferred the galaxies can be permanently evolving in 
rather closed cycles between luminous and dark states, and vice versa,
indefinitely throughout the time.

From the high proportion of matter in dark states, is inferred that the 
universe age is, at least, of a higher order of magnitude than a full 
galactic cycle, i.e., that some new galaxies have been recently born after 
``small bangs''.

In a galaxy cycle, Hydrogen can evolve between the states of gas
and LBH. The last one, after absorbing energy, would explode thus
regenerating new gas. Then the G fields of the cold bodies remaining from 
the dark period should capture the new gas that would transform them into 
new stars. This process should fix the initial states of the new luminous
periods of galaxies\cite{Vera13} \cite{Vera14} \cite{Vera 16} \cite{Vera17}.
Something similar should occur in larger systems like clusters.

\subsection{Gross evolution of luminous galaxies}

The chain of LBH explosions giving away to a galaxy should generate H gas
with a high density of randomly oriented angular momentum. Thus \textit{the
newest} luminous galaxies should have the highest volume with a rather
``spherical form''. Most of their stars would be formed from condensation of
gas around planets and planetesimals. They should have low-densities and
low temperatures (red). They should be rich in H, with low metal
contamination. Only a small fraction of their stars would be formed from gas
trapped by the strong G fields around older (dead) dwarf stars. The last 
ones should have higher temperatures. They are likely to be
consistent with the \textit{blue stragglers}.

Then the \textit{newest galaxies} should correspond with some elliptical
galaxies that, paradoxically, are currently assumed to be the 
\textit{oldest}
ones. Curiously,  the really \textit{new stars}, made up with new and clean
H, are currently called \textit{old stars}.

In a new galaxy, due to the higher average cancellation rates of angular
momentum with random orientations, compared with those of preferred
orientations, the radius of the spherical set of such stars should decrease
at faster rates compared with the more massive stars of preferred
orientations close to the galactic plane.

Then, in the long run, the elliptical galaxies should turn into \textit{disc
and spiral galaxies} with a spherical set of lower density stars whose
volume decreases with the time. Then the last galaxies are bounded to be
older (more evolved) than the first ones.

In the meanwhile the average density and temperatures of the luminous stars
should increase with the time before running out available energy. Thus the
interstellar space should become progressively contaminated with metals and
crowded with clusters of dead stars and planetesimals. The spirals with more
massive stars would run out of available energies thus becoming dark rings
around the spherical bulge.

The interstellar gas clouds should drift towards regions of bodies in lower
potentials thus regenerating new star clusters with bluer stars more
contaminated with metals. Curiously, such stars formed \textit{later on},
with older materials, are called \textit{young stars. }

The last luminous region of a dark galaxy should occur in its center, most
probably around a massive binary LBH and other neutron stars. Then the light
emitted in such low G potentials should be strongly red shifted due to GTD.
The gas captured by the neutron stars and LBHs would produce jets in their
polar regions, i.e., they may become radio sources. Then these small
luminous regions are consistent with the ``true''\textit{quasi-stellar radio
sources} (quasars) whose variations of luminosity can be compatible with
their small volumes. They should be not confused with the quiet galaxies of
high cosmological red shifts.

\subsection{The black stages of galaxies}

When galaxies run out of sources of electromagnetic radiation, they should
cool down up to very low temperatures because their LBHs should absorb
radiation coming from the dark galaxy and from the rest of the universe.
Thus, globally, a black galaxy would absorb more radiation than the one
emitted by their dark bodies.

However, during the cooling period, such galaxies can emit some radiation
coming from eventual falls of residual plasmas and other bodies into massive
neutron stars. Such falls would transform G energy into other forms of
energy that would be radiated away as cosmic and electromagnetic radiation.

Black galaxies would not collapse because, as shown above, the G field has
no energy and, therefore, it cannot\textit{\ emit gravitons}.

The energy recovering period of a black galaxy would end when the average
mass of the nucleons in some massive black hole, with respect to an external
observer, is larger than the mass of a neutron in free state. Other massive 
bodies orbiting around them can trigger the chain of LBH explosions.

\subsection{The missing mass and the low temperature background of the
universe}

In the new astrophysical scenario, statistically, all of the evolution 
stages of the galaxies should be present in the sky in the proportions 
fixed by their average evolution periods. Due to the low average density 
of radiation in the intergalactic space, the average energy-recovering 
period of the LBHs of a dark galaxy must by of a higher order of magnitude 
than the luminous period of galaxies. Then, statistically, most of the 
universe must be in the state of dark galaxy. Consequently, the dark 
galaxies should account for most of:

- The dark matter in the intergalactic space.

- The radiation backgrounds of the universe, like the low temperature cosmic
microwave background (CMB), the gamma rays and the gamma bursts coming from
the intergalactic space, and the iron content in their spectrums.

Notice that the new scenario also solves old problems in astrophysics,
which are the high differences of age or the evolution stages of bodies and 
galaxies that are relatively close to each other.

\subsection{The new model of star formation}

According to the new scenario, the first stars of a new luminous period of a
galaxy would be formed after the fall of clean gas over the cool bodies of
the previous black galaxy period. Due to the existence of such previous
``seed bodies'' this process would occur in time scales of much lower order
of magnitude than the conventional model of star formation. This would
account for the low gas content in the interstellar space of some elliptical
galaxies, and the low content of metals.

New satellites can be formed by concentration of gas and particles of common
angular momentum in rings around planets. Such bodies, after consecutive
captures of gas, may become new planets that would grow in mass until they
may become new stars and so on. Later on, a similar process should occur
after the condensation of the gas ejected from stars contaminated with
metals coming from nuclear fusion reactions. New stars of higher
temperatures should come out after condensation of gas over dead (dwarf)
stars and neutron stars.

Globular clusters can be formed from a relatively fast fall of new
gas over older clusters of cool bodies of different masses that were in
different evolution stages. This can account for all of them: the low 
interstellar gas, the low metals, the high proportion of randomly oriented 
momentum, and the high differences of masses and evolution stages of the new 
bodies that are relatively close to each other in the same star cluster or 
galaxy.

\subsection{The new role of the gravitational energy in the universe}

This new scenario also brings out important changes on the role of the G
energy in the evolution of stars and galaxies. For example,

- During a galaxy cycle, the condensation of matter, from gas state up to
LBH state, must release a net G energy of higher order of magnitude
compared with that of nuclear fusion of the original H.

- Most of such energy should be transformed in other forms of energy in
regions very close to neutron stars or LBHs.

Notice that the G binding energy of neutrons in a neutron star is higher
than the nuclear binding energy of neutrons in the atomic nuclei. Then in
principle \textit{the nuclear stripping reactions (of the Oppenheimer kind
can transform G energy into nuclear potential and kinetic energy without
neutrino emission}. This mechanism would regenerate nuclear potential energy
at the cost of the lower G potential of the captured neutrons.\cite{Vera2} 
\cite{Vera15}.

Such neutron stripping reactions may occur after the fall of plasma into
naked neutron stars. They can account for the cosmic jets and cosmic rays. 
They can also account for the fact that cosmic rays with lower proportions 
of neutrons have higher energies\cite{Vera2} \cite{Vera7} \cite{Vera15} 
\cite{Vera10}.

Due to the high proportion of the G energy that must be released during a
galactic cycle, it is reasonable that an appreciable fraction of such energy
should be released rather hidden, either steadily or after periodical
collapses, in central regions of some stars. Thus such stars should have
higher temperatures both because they would have stronger internal fields
and because they would run with two kinds of energy sources: nuclear and
gravitational energies. Then the should have lower proportions of \textit{\ 
neutrino emission}. They are consistent with the low neutrino background in 
the universe.

\section{Some conclusions}

The EEP can be used to demonstrate that some fundamental physical changes
occur to the bodies after a change of distance with respect to other bodies,
i.e., after a change of G potential. However the observers having the same
change of G potential as the bodies cannot detect such physical changes
because everything changes in the same proportion. Thus the omission of such
changes has been a common source of fundamental errors in current
gravitation and astrophysics. The main ones are the presumed energy of G
fields and the presumed invariability of the bodies after the increase of G
potential produced after a universe expansion.

According to the EEP, only an absolute kind of universe expansion can exist. 
Such expansion would not produce measurable changes after the time because 
it does not change any local ratio. Thus, statistically, the average 
universe should look the same all of the time. Then it is not obvious
whether the Hubble red shift is due to a real universe expansion or to some
unknown phenomenon occurring during the trip of the wavelets coming from
long distances.

Then the new universe age, after the cosmological period, is indefinite. 
Thus to account for the present state of the universe, it is necessary to 
accept that, statistically, matter should be evolving, rather locally in 
galaxies,  in closed cycles between the states of gas H and LBH. The 
``small bangs'' produced by the LBH explosions should fix first luminous 
stages of galaxies and clusters. Thus galaxies should also be evolving 
in closed cycles between luminous and dark states. It is simple to verify 
that all of the stages of the cyclical evolution of galaxies have been 
already detected, directly and indirectly, from astronomical observations.
Thus the universe age is, at least, larger than one galactic cycle.

Statistically, most of the matter in the universe should be in the state 
of cool galaxies that are absorbing energy from the rest of the universe.
\textit{They are consistent with the low temperature background of the 
universe and with the missing mass in the intergalactic space}. They are 
also consistent with the other kinds of radiations that come from the 
intergalactic space.

Notice that the new astrophysical scenario does not includes the true
cosmological period of formation of bodies of well-defined structure, 
after which the EEP can be applied. Then such scenario is 
not a new cosmology. Fortunately, it is independent on any cosmological 
theory. Thus this is a good scenario for astrophysical researh.

\end{document}